\documentclass[10pt, conference, letter]{IEEEtran}
\IEEEoverridecommandlockouts

\usepackage{amsmath, amsfonts, amssymb}
\usepackage{xcolor}
\usepackage{comment}
\usepackage{graphicx} 
\usepackage{multirow} 
\usepackage{hhline} 
\usepackage{subcaption} 
\usepackage{adjustbox} 
\usepackage{tikz}
\usepackage{nicematrix}
\usepackage{algorithm}
\usepackage{algpseudocode}
\usepackage{float}
\usepackage{booktabs}

\usetikzlibrary{positioning}
\usepackage{circledtext}

\title{MPX: A Unified Systolic Array for Matrix and Polynomial Multiplication}
\author{%
\IEEEauthorblockN{George Alexakis}
\IEEEauthorblockA{Electrical and Computer Engineering\\ 
Democritus University of Thrace, Xanthi, Greece}
\and
\IEEEauthorblockN{Dimitrios Schoinianakis}
\IEEEauthorblockA{Nokia Bell Labs\\  
Athens, Greece}
\and
\IEEEauthorblockN{Giorgos Dimitrakopoulos}
\IEEEauthorblockA{Electrical and Computer Engineering\\ 
Democritus University of Thrace, Xanthi, Greece}
}

\begin{document}

\maketitle

\begin{abstract}
Polynomial multiplication is a fundamental kernel in Fully Homomorphic Encryption (FHE) and post-quantum cryptography (PQC) and is commonly accelerated through Number Theoretic Transforms (NTTs). To avoid the cost of designing dedicated cryptographic accelerators, recent efforts have mapped NTT computations onto existing systolic matrix engines, enabling the reuse of AI hardware for cryptographic workloads.
In this work, we take the opposite approach. We observe that the wavefront dataflow of systolic arrays naturally aligns with the accumulation pattern of polynomial multiplication and leverage this correspondence to design MPX, a dual-mode systolic array that supports both matrix multiplication and direct polynomial multiplication within the same hardware fabric.
Experimental results show that extending a conventional systolic array with this dual-mode capability requires only 20\% additional area and introduces negligible power overhead during matrix-multiplication execution. In polynomial-multiplication mode, MPX achieves more than 1.2$\times$ lower latency compared to NTT-based polynomial multiplication on systolic matrix engines.
\end{abstract}

\section{Introduction}

Polynomial multiplication is a fundamental operation in modern lattice-based cryptography, forming the computational backbone of both Fully Homomorphic Encryption (FHE) and many post-quantum cryptographic (PQC) schemes~\cite{gong2024practical, zeng2024implementation}. Despite the strong security and functionality guarantees offered by these systems, their practical deployment remains constrained by the high computational cost. While specialized accelerators have been proposed to address this challenge~\cite{zhang2024sok, daksha2026fhecore}, designing dedicated hardware for rapidly evolving FHE and PQC algorithms is costly and difficult to scale.

An increasingly attractive alternative is to leverage existing AI hardware~\cite{tpu}. Modern accelerators built around systolic arrays (SAs) provide massive parallelism and energy efficiency, motivating a growing body of work that maps cryptographic workloads onto AI accelerators~\cite{fan2023tensorfhe,sun2025tensorfhe+,tong2026leveraging}. The challenge is that FHE and PQC computations are organized around polynomial arithmetic, whereas SAs are designed to efficiently execute matrix operations.

Most prior approaches bridge this gap through the Number Theoretic Transform (NTT)~\cite{harvey2014faster}. By expressing polynomial multiplication as a sequence of NTTs and pointwise multiplications, these methods expose regular computations that can be executed efficiently on matrix engines~\cite{tong2026leveraging}. Effectively, this approach adapts polynomial arithmetic to the currently available capabilities of matrix hardware.

In this work, we explore the opposite direction. Rather than transforming polynomial multiplication to fit existing matrix engines, we adapt the matrix engine itself to directly support polynomial multiplication. We observe that the wavefront propagation pattern of systolic arrays naturally aligns with the accumulation pattern of polynomial multiplication, enabling polynomial arithmetic to execute directly within the SA.

Guided by this observation, we design MPX, a dual-mode systolic architecture that supports both matrix multiplication and polynomial multiplication within the same hardware structure. MPX preserves the hardware-reuse benefits of AI matrix engines while eliminating the need to route polynomial multiplication through an NTT-centric execution flow.
The contributions of this work can be summarized as follows:
\begin{itemize}
\item We present MPX, the first SA architecture, to the best of our knowledge, that directly maps polynomial multiplication onto the existing multiply-accumulate processing elements of a SA.

\item The introduced polynomial-multiplication dataflow supports arbitrary sequences of back-to-back polynomial multiplications, enabling the execution of large polynomial multiplications on SAs with fixed dimensions.

\item Experimental results show that MPX extends a conventional SA with native polynomial-multiplication support at 20\% area overhead, while achieving more than 1.2$\times$ speedup compared to NTT-based polynomial multiplication approaches.
\end{itemize}

\section{Background and Related Work}
FHE and PQC workloads are dominated by polynomial arithmetic, which is commonly accelerated through the Number Theoretic Transform (NTT). NTT converts polynomial convolution into element-wise multiplication in the transform domain, making it a key building block for efficient polynomial multiplication~\cite{harvey2014faster}.

Recent work has increasingly explored the reuse of AI accelerators for NTT execution. For example, TensorFHE~\cite{fan2023tensorfhe}, TensorFHE+~\cite{sun2025tensorfhe+}, and CROSS~\cite{tong2026leveraging} demonstrate that cryptographic kernels can be reformulated as dense linear-algebra operations and mapped onto tensor cores and SAs. This is typically achieved using the 4-step NTT algorithm~\cite{bailey1990ffts}, which decomposes large transforms into a sequence of matrix-friendly operations.

The 4-step NTT first reshapes the input vector into a matrix and then performs column-wise transforms, twiddle-factor multiplication, a matrix transpose, and row-wise transforms. The transform stages are expressed as matrix multiplications, while the remaining stages consist of element-wise operations and data reorganization.

Although this approach enables polynomial arithmetic to leverage the computational throughput of existing matrix engines, it also introduces additional transpositions, buffering, synchronization, and data movement overheads.

\begin{figure}[t]
\centering
\includegraphics[width=0.8\columnwidth]{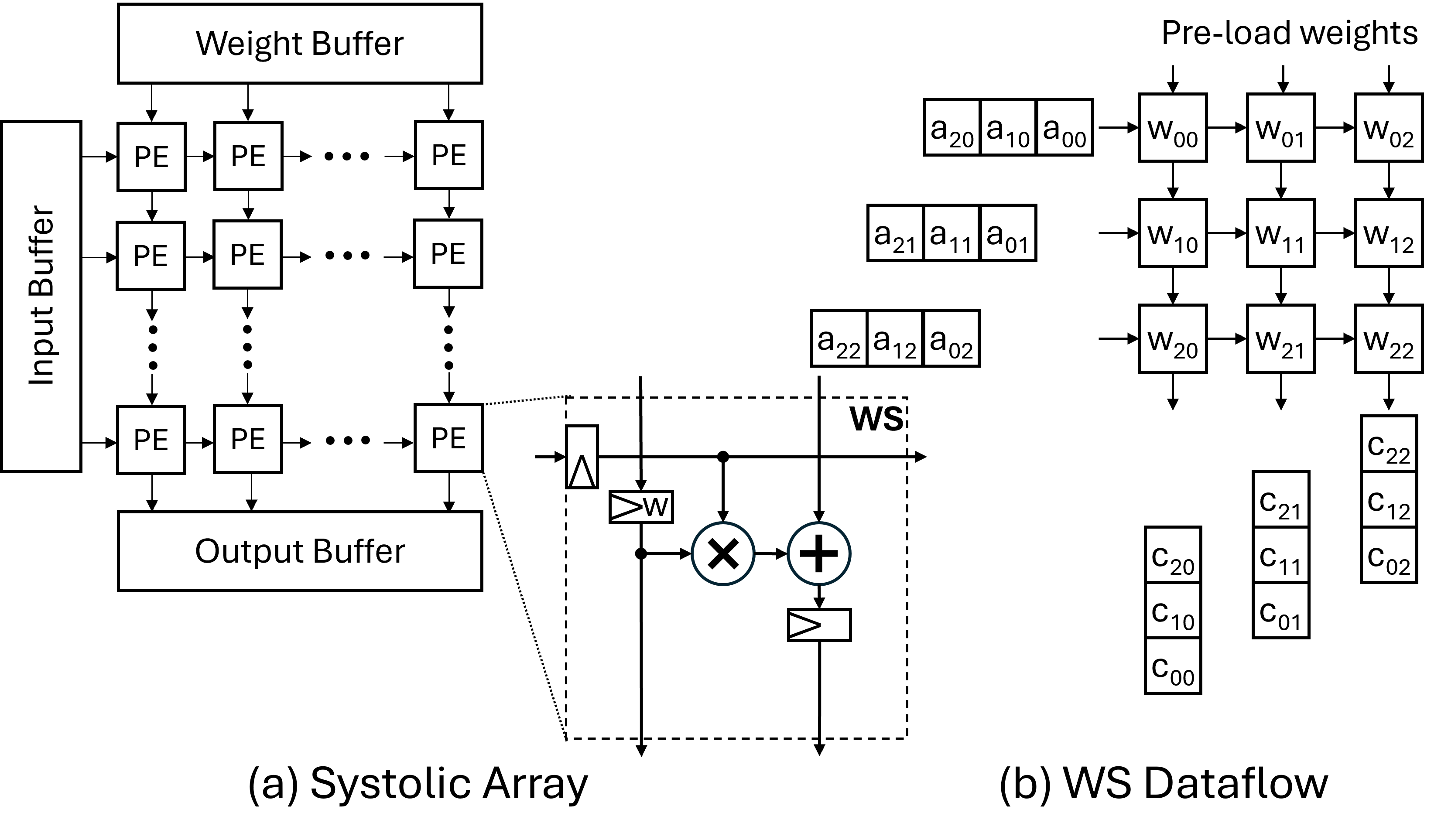}
\caption{The micro-architecture of a typical SA, and a high-level overview of the weight-stationary (WS) dataflow.}
\label{f:sa-baseline}
\end{figure}

A typical SA consists of a 2D array of processing elements (PEs), as shown in Fig.~\ref{f:sa-baseline}(a). Each PE contains a multiplier, an adder, and the registers required to sustain pipelined data movement through the array. Input operands are streamed from local memory banks located along the North and West edges, while outputs are collected along the South edge.
The organization of computation within the SA is determined by its dataflow. In the widely adopted weight-stationary dataflow~\cite{raj2025scale}, weights are preloaded into the array while input is streamed from the West side, as illustrated in Fig.~\ref{f:sa-baseline}(b). The matrix product is computed through a wavefront of multiply-accumulate operations that propagates diagonally across the fabric.

\begin{figure}[t]
\centering
\begin{subfigure}{\linewidth}
\centering
\includegraphics[width=0.78\columnwidth]{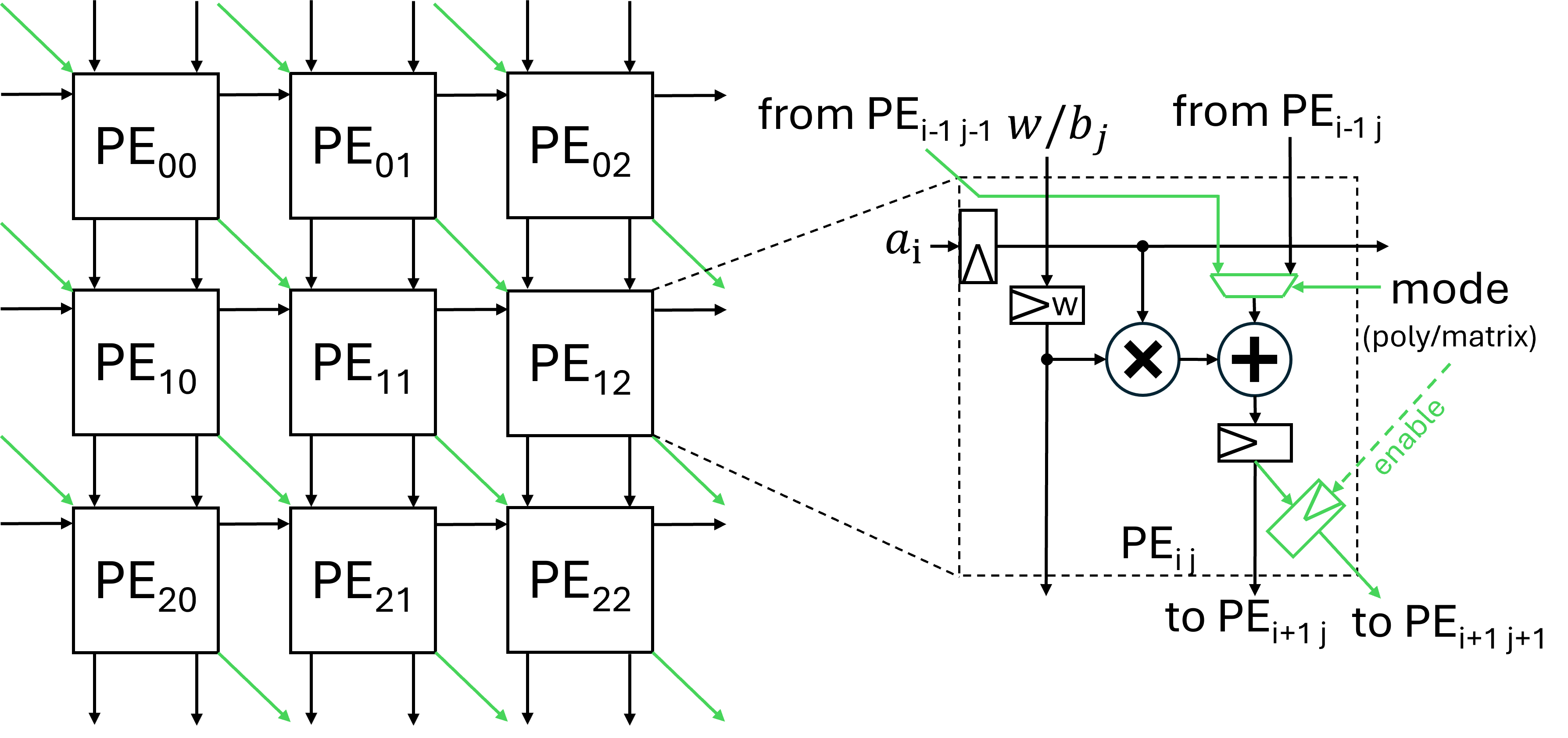}
\caption{MPX}
\label{fig:proposed_dual}
\end{subfigure}
    \hfill
    \begin{subfigure}{\linewidth}
        \centering
        \includegraphics[width=0.78\columnwidth]{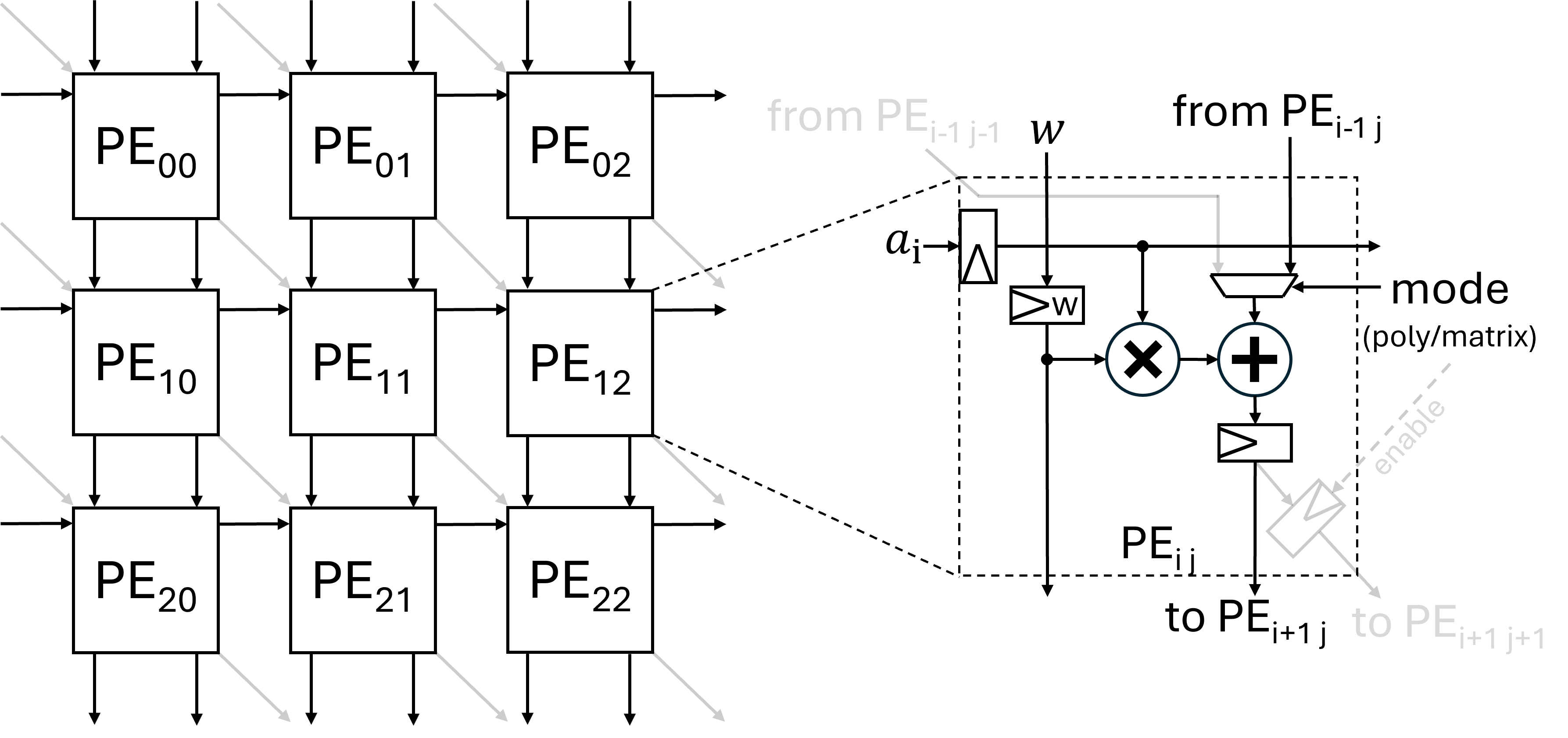}
        \caption{Matrix Multiplication mode}
        \label{fig:proposed_matrix}
    \end{subfigure}
    \hfill
    \begin{subfigure}{\linewidth}
        \centering
        \includegraphics[width=0.78\columnwidth]{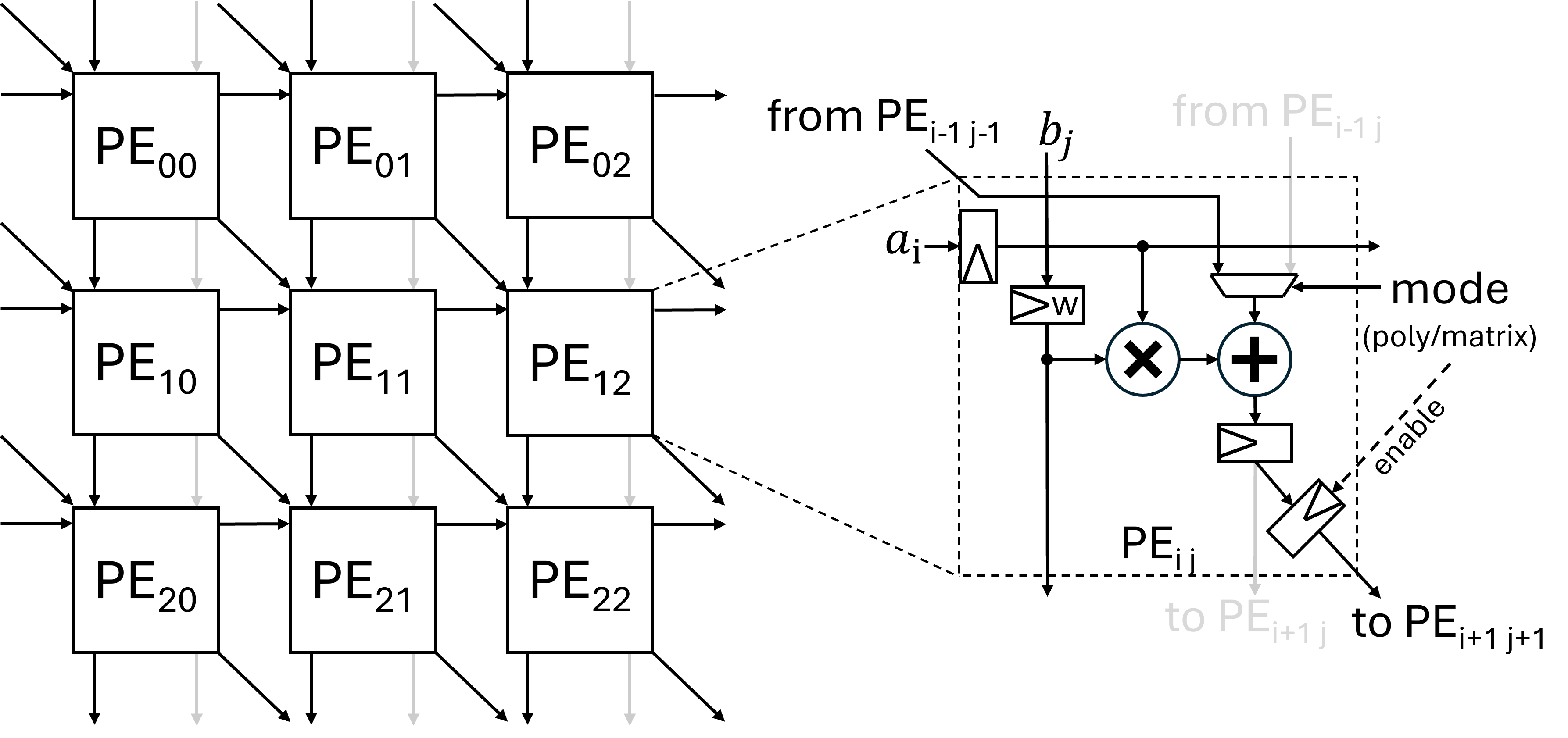}
        \caption{Polynomial Multiplication mode}
        \label{fig:proposed_poly}
    \end{subfigure}
    \caption{(a) MPX PE supporting two modes of operation. The corresponding dataflows are highlighted separately in (b) weight-stationary matrix-multiplication mode and (c) polynomial-multiplication mode.}
    \label{fig:proposed}
\end{figure}

\begin{figure*}[t!]
    \centering
    \includegraphics[width=0.7\textwidth]{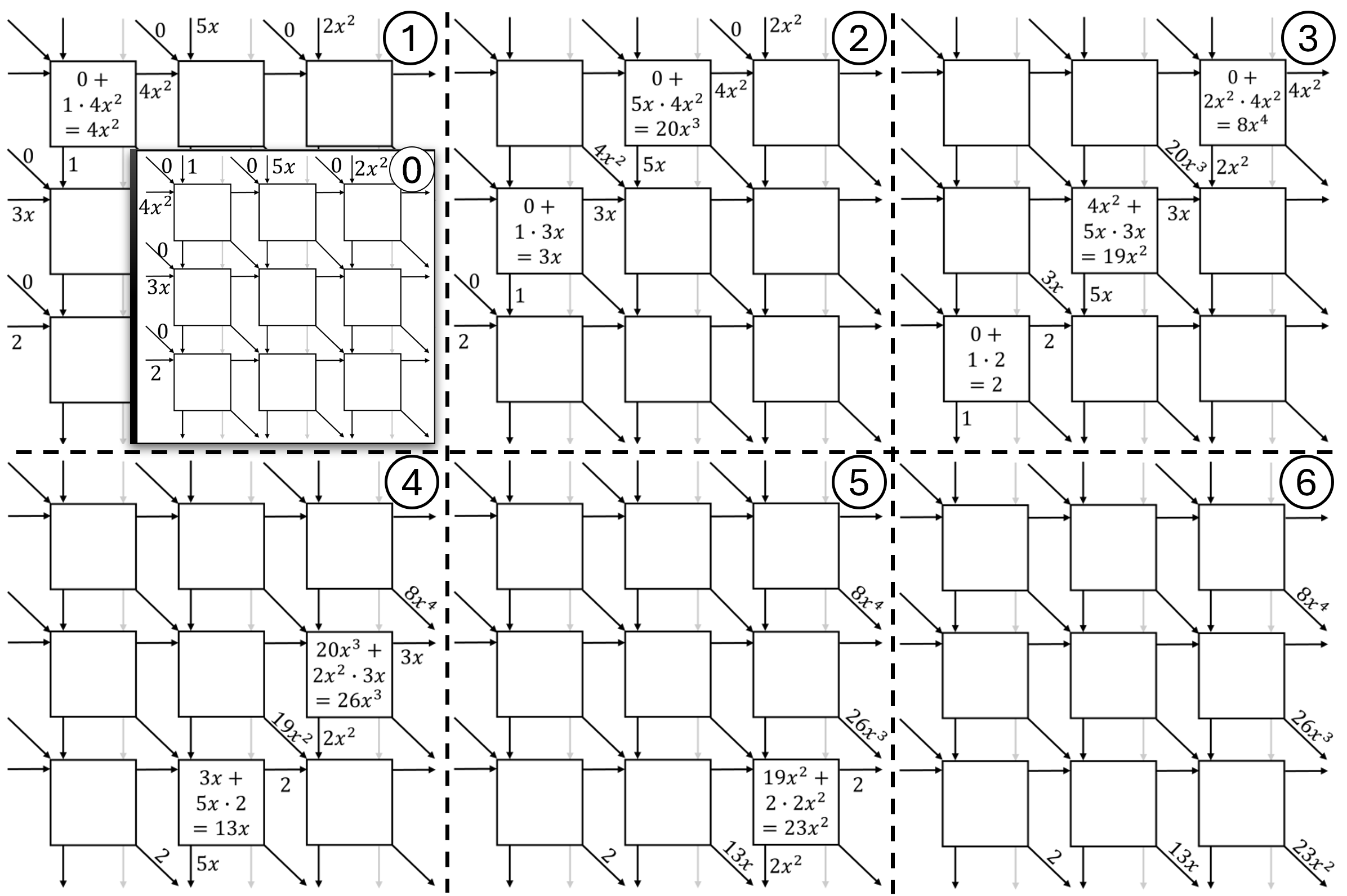}
    \caption{Running example of polynomial multiplication on MPX.}
    \label{fig:example}
\vspace{-12pt}
\end{figure*}

\section{Unified Systolic Array for Matrix and Polynomial Multiplications}
In this work, instead of adapting polynomial multiplication to existing matrix engines, we extend the SA itself to directly support polynomial multiplication. The objective is to preserve the efficiency of matrix-multiplication accelerators while enabling polynomial multiplication through minimal architectural modifications.

The key observation behind MPX is that polynomial multiplication naturally maps to the wavefront execution model of SAs.
Given $A(x) = \sum_{i=0}^{n-1} a_i x^i$ and $B(x) = \sum_{j=0}^{m-1} b_j x^j$, their product $C(x)=\sum_{k=0}^{n+m-2} c_k x^k$ has coefficients $c_k$ that are defined by the sum of all cross-products with index sum $i+j$ equal to $k$, i.e., $c_k = \sum_{i+j=k} a_i b_j$.
%
As a result, the products contributing to a given coefficient lie along a common diagonal in the 2D $(i,j)$ index space, closely matching the wavefront propagation already present in SAs.

\subsection{Dual-mode Systolic Array}
Leveraging this correspondence, Fig.~\ref{fig:proposed} presents MPX, a dual-mode systolic architecture that supports both matrix multiplication and polynomial multiplication within the same SA. Relative to the baseline SA, MPX introduces a second accumulation path for polynomial multiplication. In matrix mode, inputs propagate horizontally while partial sums follow the conventional vertical reduction path. In polynomial mode, the two input polynomials are streamed from orthogonal directions (North and West), with partial sums forwarded diagonally across the array. To support this operation, the vertical weight-loading path used during matrix multiplication is reused to stream the second polynomial, allowing products contributing to the same output coefficient to be accumulated as they traverse the systolic fabric.

Correct operation requires the diagonally-propagated partial sums to remain synchronized with the incoming operands. To maintain this alignment, MPX introduces a single pipeline register along the diagonal polynomial accumulation path. This lightweight modification preserves the critical path of the SA, while ensuring that operands and partial sums arrive at each PE at the appropriate cycle for accumulation.

By supporting both accumulation patterns within the same hardware structure, MPX enables direct polynomial multiplication while remaining fully compatible with conventional matrix workloads.

To illustrate the polynomial multiplication dataflow, Fig.~\ref{fig:example} shows the multiplication of 
$2+3x+4x^2$ and $1+5x+2x^2$ in MPX.
As coefficients stream into the array from orthogonal directions, each PE generates a local partial product. In polynomial mode, these partial products are accumulated along the diagonals as they traverse the array. Consequently, all products contributing to the same output coefficient meet and are added within the SA.


\subsection{Scaling to Larger Polynomial Multiplications}
To support polynomials exceeding the array's physical dimensions, operands are decomposed into $L$-coefficient blocks. A polynomial $A(x)$ is partitioned as $A(x)=\sum_i A_i(x)x^{iL}$. The product $A(x)B(x)$ is then reconstructed by computing all pairwise convolutions $A_i(x)B_j(x)$ natively on the hardware and accumulating them with the corresponding $x^{(i+j)L}$ shift.

\begin{figure}[t]
\centering
\includegraphics[width=0.78\columnwidth]{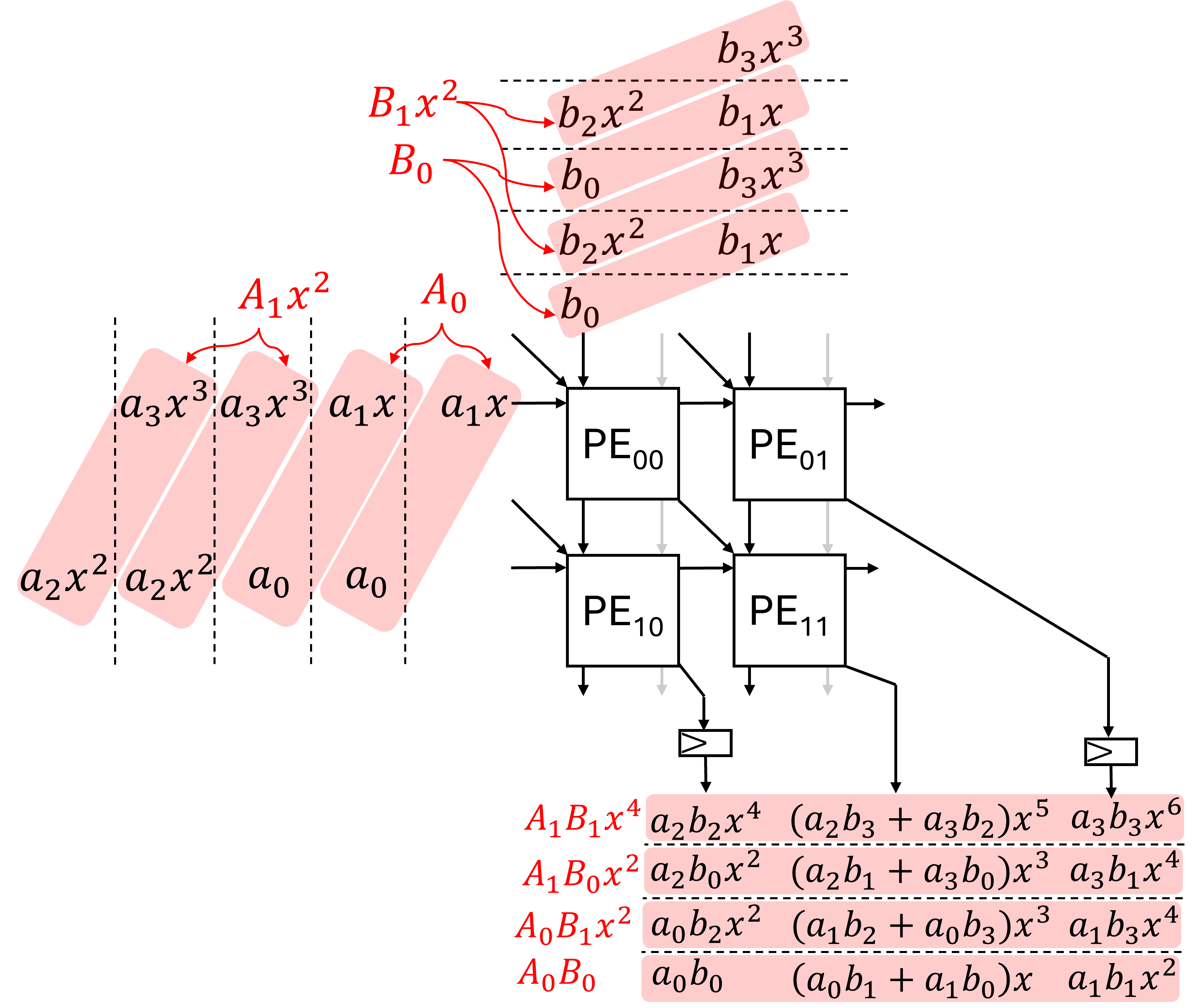}
\caption{Back-to-back execution of sub-polynomial multiplications on a 2$\times$2 MPX.}
\label{f:sub-polys}
\vspace{-10pt}
\end{figure}

For example, multiplying degree-4 polynomials on an array with block length $L=2$ partitions the inputs into $A(x)=A_0(x)+A_1(x)x^2$ and $B(x)=B_0(x)+B_1(x)x^2$, where $A_0(x)=a_0+a_1x$ and $A_1(x)=a_2+a_3x$ (with an analogous decomposition for $B$). The multiplication expands to
$$
A(x)B(x)=A_0B_0+(A_0B_1+A_1B_0)x^2+A_1B_1x^4.
$$
This reduces the larger multiplication to four independent, hardware-native polynomial multiplications. As illustrated in Fig.~\ref{f:sub-polys}, the sub-polynomials are streamed into MPX back-to-back without introducing idle cycles between successive computations, allowing MPX to remain continuously utilized. To reconstruct the final polynomial, the resulting products are accumulated in vector units outside MPX.

\section{Evaluation}
We evaluate MPX from two complementary perspectives. First, we quantify the hardware overhead of extending a conventional SA with native polynomial-multiplication support. Second, we compare the latency of direct polynomial multiplication on MPX against state-of-the-art approaches that perform polynomial multiplication through NTTs mapped onto existing AI accelerators.

\subsection{Hardware Cost of Dual-Mode Operation}
To quantify the hardware cost of MPX, we compare the proposed architecture against a conventional SA of identical dimensions. The additional hardware consists of two components: (i) mode-controlled routing and multiplexing within each PE to support diagonal accumulation, and (ii) an additional pipeline register in the polynomial datapath to maintain operand alignment. The PEs employ 8-bit integer multipliers and 32-bit adders, matching the arithmetic precision of industrial systolic arrays~\cite{tpu, amx}.

Both architectures were implemented using the OpenROAD physical-design flow and the ASAP7 technology library. All designs were synthesized and placed-and-routed at a target clock frequency of 1 GHz.

\begin{table}[t]
\centering
\caption{Post-layout area and power comparison between the baseline SA and MPX. For MPX power is reported separately per mode of operation.}
\setlength{\tabcolsep}{2pt}
\begin{tabular}{@{}c|cc|ccc@{}}\toprule
\multirow{3}{*}{\shortstack{SA \\ Size}}    & \multicolumn{2}{c|}{Area (mm\textsuperscript{2})} & \multicolumn{3}{c}{Power (W)}                  \\
  & \multirow{2}{*}{Baseline}  & \multirow{2}{*}{MPX} & \multirow{2}{*}{Baseline}  & MPX & MPX             \\
 &  &                     &   & Matrix Mode  & Poly mode                 \\
\midrule
16x16 & 0.017     & 0.020   & 0.101  & 0.104 & 0.146  \\
32x32 & 0.066     & 0.079   & 0.405  & 0.417 & 0.587    \\
64x64 & 0.263     & 0.317   & 1.632  & 1.682 & 2.367     \\
\bottomrule
\end{tabular}
\label{tab:areapower}
\end{table}

\begin{figure}[t]
    \centering
    \includegraphics[width=0.7\columnwidth]{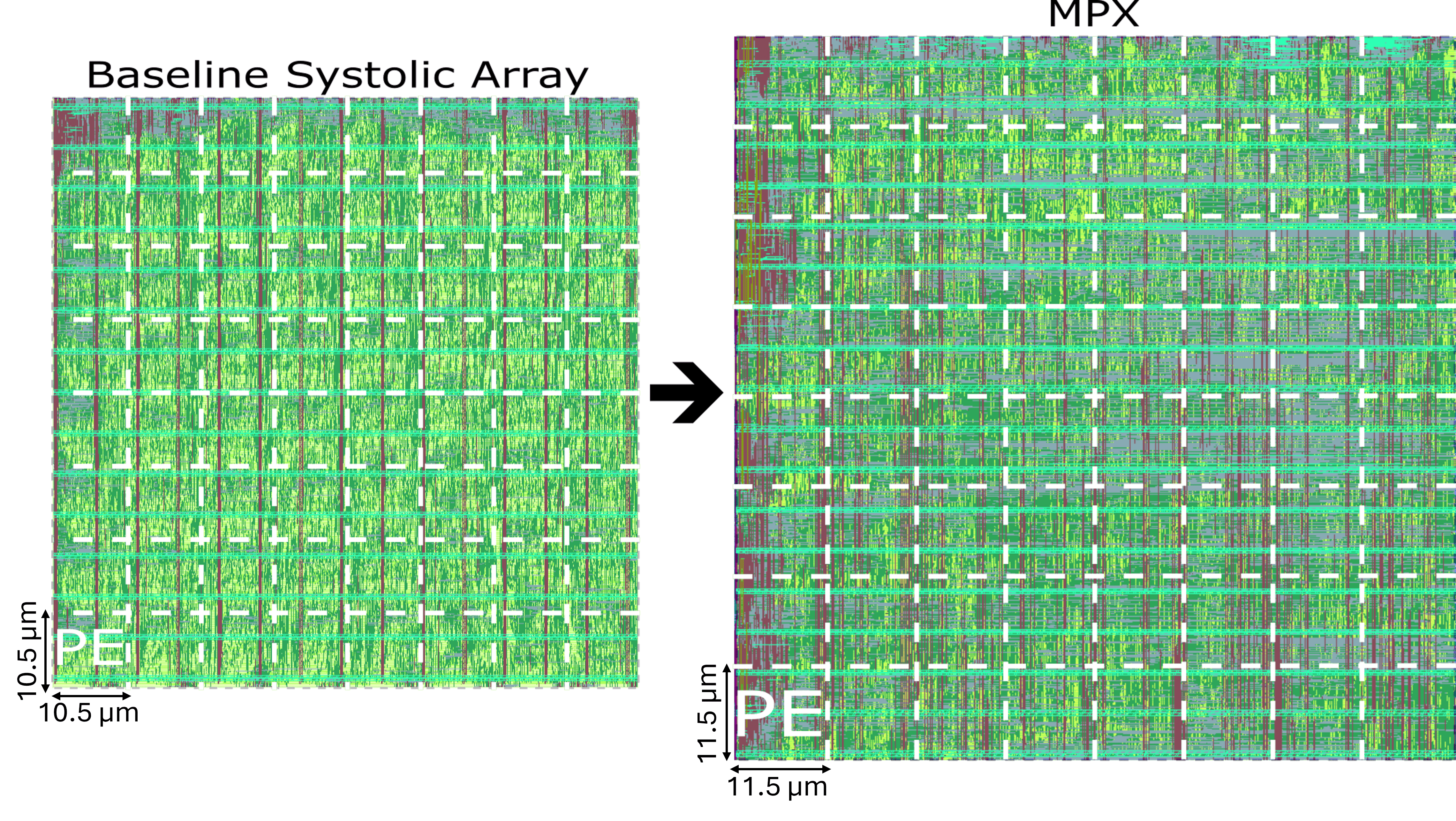}
    \caption{Physical layout of 8$\times$8 instances of the baseline SA and the proposed MPX. }
    \label{fig:layout}
\end{figure}

Table~\ref{tab:areapower} summarizes the post-layout area and power results across multiple SA sizes, while Fig.~\ref{fig:layout} illustrates the physical layouts of a baseline $8\times8$ SA and an MPX design of the same size. The additional hardware required to support both matrix- and polynomial-multiplication modes increases area by approximately 20\%, with similar overhead observed across all evaluated array dimensions.

Power is reported separately for each operating mode. In matrix-multiplication mode, MPX incurs only 3\% additional power consumption relative to the baseline SA, primarily due to the added 32-bit multiplexing logic. The diagonal registers and routing associated with polynomial execution remain clock-gated and therefore do not contribute to dynamic power consumption.
In polynomial-multiplication mode, power consumption increases relative to matrix mode due to the activation of the diagonal accumulation path, including the additional 32-bit registers and interconnects required for diagonal data propagation.

Overall, MPX achieves its goal of supporting both matrix and polynomial multiplication within the same SA while introducing minimal power overhead for conventional matrix-multiplication workloads.

\subsection{Performance Evaluation}

\begin{table}[t]
\centering
\caption{Number of cycles for computing polynomial multiplication either directly in MPX or through matrix-multiplication-based NTTs.}
\begin{tabular}{@{}ccccc@{}}
\toprule
Polynomial & SA & \multirow{2}{*}{NTT-based} & \multirow{2}{*}{MPX} & \multirow{2}{*}{Speedup} \\ 
Degree & Size & & & \\ \midrule
\multirow{3}{*}{128} & 16x16    & 567      & 159      & 3.57    \\
                     & 32x32    & 855      & 95       & 9.00    \\
                     & 64x64    & 1431     & 135      & 10.60   \\\midrule
\multirow{3}{*}{256} & 16x16    & 766      & 543      & 1.41    \\
                     & 32x32    & 1054     & 191      & 5.52    \\
                     & 64x64    & 1630     & 159      & 10.25   \\\midrule
\multirow{3}{*}{512} & 16x16    & 2500     & 2079     & 1.20    \\
                     & 32x32    & 1546     & 575      & 2.69    \\
                     & 64x64    & 2122     & 255      & 8.32    \\ \bottomrule
\end{tabular}
\label{tab:cycles}
\end{table}

Table~\ref{tab:cycles} compares MPX against a conventional SA executing polynomial multiplication through a four-step NTT decomposition for various polynomial degrees and SA sizes.
All configurations were evaluated using SCALE-Sim v3~\cite{raj2025scale} after modeling MPX. Operations that do not execute inside the SA, such as element-wise multiplications, or sub-polynomial accumulations, were modeled using SCALE-Sim's SIMD backend configured with 16 lanes to approximate an AVX-512-class vector unit for 32-bit element vectors.

MPX consistently outperforms the NTT-based baseline across the evaluated configurations. The advantage becomes more pronounced as the SA dimensions increase, since direct polynomial multiplication is able to exploit the available spatial parallelism more effectively than the matrix-oriented decomposition imposed by the four-step NTT. As polynomial sizes grow, the relative benefit decreases because both approaches become increasingly dominated by the multiply-add arithmetic intensity. 

Importantly, configurations that favor NTT-based execution do not diminish the utility of MPX. Since MPX retains full matrix-multiplication functionality, NTT computations can be executed on the same hardware when they are the preferred algorithmic choice. MPX provides the flexibility to select between direct polynomial multiplication or NTT-based execution according to the characteristics of the target workload.

\section{Conclusions}
This work presented MPX, a dual-mode SA that extends conventional matrix-multiplication engines with native support for polynomial multiplication. Unlike prior approaches that accelerate polynomial arithmetic through NTT computations mapped onto matrix engines, MPX exploits the natural correspondence between polynomial multiplication and systolic wavefront execution to perform polynomial multiplication directly within the SA. With lightweight modifications to the PEs and dataflow, MPX supports both matrix and polynomial multiplication on the same hardware substrate, enabling efficient reuse across AI and FHE/PQC workloads.
\bibliographystyle{IEEEtran}
\bibliography{refs}

@article{harvey2014faster,
  title={Faster arithmetic for number-theoretic transforms},
  author={Harvey, David},
  journal={Journal of Symbolic Computation},
  volume={60},
  pages={113--119},
  year={2014},
  publisher={Elsevier}
}

@inproceedings{tong2026leveraging,
  title={Leveraging {ASIC AI} Chips for Homomorphic Encryption},
  author={Tong, Jianming and others},
  booktitle={IEEE Inter. Symp. on High Perf. Comp. Arch. (HPCA)},
  year={2026},
  organization={}
}

@inproceedings{fan2023tensorfhe,
  title={Tensorfhe: Achieving practical computation on encrypted data using gpgpu},
  author={Fan, Shengyu and others},
  booktitle={IEEE Inter. Symp. on High-Perf. Comp. Arch. (HPCA)},
  pages={922--934},
  year={2023},
  organization={}
}

@article{sun2025tensorfhe+,
  title={Tensorfhe+: Fully homomorphic encryption acceleration based on linear algebra},
  author={Sun, Yintai and others},
  journal={IEEE Trans. on Computers},
  year={2025},
}

@inproceedings{raj2025scale,
  title={SCALE-Sim v3: A modular cycle-accurate systolic accelerator simulator for end-to-end system analysis},
  author={Raj, Ritik and others},
  booktitle={IEEE Intern. Symp. on Perf.e Analysis of Systems and Software (ISPASS)},
  pages={186--200},
  year={2025},
}

@INPROCEEDINGS{tpu,
  author={Jouppi, Norman P. and others},
  booktitle={IEEE Inter. Symp. on Comp. Arch. (ISCA)}, 
  title={Ten Lessons From Three Generations Shaped {Google’s TPUv4i} : Industrial Product}, 
  year={2021},
  volume={},
  number={},
}

@techreport{amx,
  author      = "Intel",
  title       = "Architecture Instruction Set Extensions and Future Features Programming Reference",
  institution = "Intel",
  year        = "2026"
}

@article{bailey1990ffts,
  title={{FFTs} in external or hierarchical memory},
  author={Bailey, David H},
  journal={The journal of Supercomputing},
  volume={4},
  number={1},
  pages={23--35},
  year={1990},
  publisher={Springer}
}

@article{zhang2024sok,
  title={Sok: Fully homomorphic encryption accelerators},
  author={Zhang, Junxue and others},
  journal={ACM Computing Surveys},
  volume={56},
  number={12},
  pages={1--32},
  year={2024},
  publisher={ACM New York, NY}
}

@article{gong2024practical,
  title={Practical solutions in fully homomorphic encryption: a survey analyzing existing acceleration methods},
  author={Gong, Yanwei and others},
  journal={Cybersecurity},
  volume={7},
  number={1},
  pages={5},
  year={2024},
  publisher={Springer}
}

@article{zeng2024implementation,
  title={The implementation of polynomial multiplication for lattice-based cryptography: A survey},
  author={Zeng, Chenkai and others},
  journal={Journal of Information Security and Applications},
  volume={83},
  pages={103782},
  year={2024},
  publisher={Elsevier}
}

@article{daksha2026fhecore,
  title={FHECore: Rethinking GPU Microarchitecture for Fully Homomorphic Encryption},
  author={Daksha, Lohit and others},
  journal={arXiv preprint arXiv:2602.22229},
  year={2026}
}

\end{document}